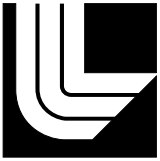

LAWRENCE
LIVERMORE
NATIONAL
LABORATORY

# Nuclear Data to Reduce Uncertainties in Reactor Antineutrino Measurements

C. Romano, N. Bowden, A. Conant, B. Goldblum, P. Huber, J. Link, B. Littlejohn, H. P. Mumm, J. Ochoa, S. Prasad, C. Riddle, A. Sonzogni, W. Wieselquist

December 8, 2021

# Nuclear Data to Reduce Uncertainties in Reactor Antineutrino Measurements

*Summary Report of the Workshop on Nuclear Data for Reactor Antineutrino Measurements (WoNDRAM)*

Catherine Romano, Nathaniel Bowden, Andrew Conant, Bethany Goldblum, Patrick Huber, Jonathan Link, Bryce Littlejohn, Pieter Mumm, Juan Pedro Ochoa-Ricoux, Shikha Prasad, Catherine Riddle, Alejandro Sonzogni, William Wieselquist

# Nuclear Data to Reduce Uncertainties in Reactor Antineutrino Measurements
*Summary Report of the*
*Workshop on Nuclear Data for Reactor Antineutrino Measurements (WoNDRAM)*

Catherine Romano[1], Nathaniel Bowden[2], Andrew Conant[3], Bethany Goldblum[4], Patrick Huber[5], Jonathan Link[5], Bryce Littlejohn[6], Pieter Mumm[7], Juan Pedro Ochoa-Ricoux[8], Shikha Prasad[9], Catherine Riddle[10], Alejandro Sonzogni[11], William Wieselquist[12]

[1]IB3 Global Solutions, Oak Ridge, TN 37830, USA
[2]Nuclear and Chemical Sciences Division, Lawrence Livermore National Laboratory, Livermore, CA, 94550, USA
[3]Nuclear Nonproliferation Division, Oak Ridge National Laboratory, Oak Ridge, TN 37831, USA
[4]Nuclear Science Division, Lawrence Berkeley National Laboratory, Berkeley, CA 94720, USA
[5]Virginia Tech, Department of Physics, Blacksburg, VA 24061, USA
[6]Illinois Institute of Technology, Department of Physics, Chicago, IL 60616, USA
[7]National Institute of Standards and Technology, Neutron Physics Group, Gaithersburg, MD 20899, USA
[8]University of California at Irvine, Department of Physics and Astronomy, Irvine, CA 92617, USA
[9]Texas A&M University, Department of Nuclear Engineering, College Station, TX 77843, USA
[10]Nuclear Science & Technology Directorate, Idaho National Laboratory, Idaho Falls, ID 83415, USA
[11]Nuclear Science and Technology Department, Brookhaven National Laboratory, Upton, NY 11973, USA
[12]Nuclear Energy and Fuel Cycle Division, Oak Ridge National Laboratory, Oak Ridge, TN 37831, USA

The support for WoNDRAM was provided by the NDIAWG. This work has been performed under the auspices of the Brookhaven National Laboratory under Contract No. DE- AC02-98CH10886 with Brookhaven Science Associates, LLC., Lawrence Berkeley National Laboratory under contract No. DE-AC02-05CH1123, Lawrence Livermore National Laboratory under contract DE-AC52-07NA27344, Oak Ridge National Laboratory under Contract No. DE-AC05-00OR22725, Idaho National Laboratory under Contract No. DE-AC07-05ID14517, and the U.S. Department of Energy National Nuclear Security Administration via the Nuclear Science and Security Consortium under Award No. DE-NA0003180.
LLNL-TR-829851

**TABLE OF CONTENTS**

# ABSTRACT

The large quantities of antineutrinos produced through the decay of fission fragments in nuclear reactors provide an opportunity to study the properties of these particles and investigate their use in reactor monitoring. The reactor antineutrino spectra are measured using specialized, large area detectors that detect antineutrinos through inverse beta decay, electron elastic scattering, or coherent elastic neutrino nucleus scattering; although, inverse beta decay is the only demonstrated method so far. Reactor monitoring takes advantage of the differences in the antineutrino yield and spectra resulting from uranium and plutonium fission providing an opportunity to estimate the fissile material composition in the reactor. Recent experiments reveal a deviation between the measured and calculated antineutrino flux and spectra (the reactor anomaly) indicating either the existence of yet undiscovered neutrino physics, uncertainties in the reactor source term calculation, incorrect nuclear data, or a combination of all three.

To address the nuclear data that impact the antineutrino spectrum calculations and measurements, an international group of over 180 experts in antineutrino physics, reactor analysis, detector development, and nuclear data came together during the Workshop on Nuclear Data for Reactor Antineutrino Measurements (WoNDRAM) to discuss nuclear data needs and achieve concordance on a set of recommended priorities for nuclear data improvements. Three topical sessions focused on the reactor source term, the antineutrino spectrum, and the detector response, provided a forum to gain consensus amongst the participants on the most important data improvements to address two goals: 1) understand the reactor anomaly and 2) improve the ability to monitor reactors using antineutrinos. This report summarizes the outcomes of the workshop discussions and the recommendations for nuclear data efforts that reduce reactor antineutrino measurement uncertainties.

# 1. INTRODUCTION

Ever since the discovery of neutrinos from the Savannah River Plant, nuclear reactors have been at the forefront of neutrino research. The KamLAND experiment provided the first indisputable observation of neutrino oscillations using terrestrial anti-neutrinos produced by Japanese nuclear reactors. Recently, the Daya Bay, Double Chooz, and RENO experiments searched for neutrino oscillations at short (< 2 km) distances from reactors and successfully provided a precise determination of the $\theta_{13}$ mixing angle. These and other experiments have recently enabled to test predictions of the antineutrino source term with unprecedented precision.

Reactor antineutrino flux predictions are calculated using one of two methods: the conversion method or the summation method. In the former, the measured beta spectra from the fission of uranium and plutonium isotopes are fit with virtual beta decay branches to obtain the corresponding antineutrino spectra. In the latter, the beta spectra from the decay of each fission fragment produced during fission are summed. The Huber-Mueller (HM) model, which was developed in 2011, combines the conversion calculation of Huber for the fissile isotopes $^{235}$U, $^{239}$Pu, and $^{241}$Pu with the summation calculation of Mueller et al. for $^{238}$U. When compared to reactor measurements, it is found to overpredict the total flux by ~5%, a discrepancy commonly referred to as the "reactor antineutrino anomaly." Likewise, when the measured spectra are compared to the expectation from the HM or summation methods, a significant excess around 5 MeV is consistently observed by most experiments. The reactor antineutrino anomaly could be caused by exotic neutrino oscillations with a fourth "sterile" neutrino. However, the existence of both anomalies makes it likely that issues with the predictions are at play, which could include inaccurate nuclear data, unaccounted systematic uncertainties, or others.

Recent antineutrino measurements are targeted at learning more about the source of these anomalies. Experiments like Daya Bay and RENO can extricate from their data the yield and spectra of antineutrinos produced from the fission of the two dominant isotopes, $^{235}$U and $^{239}$Pu, and compare them with their predictions. Likewise, experiments like PROSPECT and STEREO are located near HEU reactors and thus make a pure measurement of antineutrinos from $^{235}$U fission. The data at hand are consistent with the 5 MeV excess occurring in the spectra of both isotopes and suggest that an overestimation of the $^{235}$U fission yield in the HM model is the primary contributor to the reactor antineutrino anomaly, weakening the sterile-neutrino interpretation. New data are expected from these and other experiments like JUNO-TAO that will shed additional light.

Several types of detectors are used for these measurements, and detailed models of the detector response can be extremely important. The most common type of detector uses Gd or Li doped liquid scintillator, or plastic scintillator to observe the positron and neutron emitted when an antineutrino interacts with a proton via inverse beta decay. The positron is detected first through its loss of kinetic energy and the annihilation photons. The neutron slows down and is captured by either Gd or Li, producing a characteristic signal in the detector.

## 1.1. NUCLEAR DATA REQUIREMENTS

Nuclear data are the particle interaction and decay data that are used in high-fidelity models and theoretical calculations for nearly every nuclear application. For reactor antineutrino measurements, the data required begins with the calculation of the source term from the reactor. Some of the variables involved include neutron capture and fission cross sections, independent fission yields, short-lived isotope decay data, and spontaneous fission data. Other data such as neutron scattering impact the neutron spectrum within the reactor which, in turn, can impact the neutron induced fission yields. The fission product decay data are a critical piece of data for the calculation of the antineutrino spectrum. Finally, models of the detector response require neutron scattering and capture cross sections with low uncertainty as well as quenching factors.

Several projects are currently underway to enhance the required nuclear data, including measurements of fission yields, a new evaluation of fission yields as a function of incident neutron energy, and short-lived fission product decay measurements. These data impact many applications which makes them a high priority for improvements.

## 1.2. THE NUCLEAR DATA WORKING GROUP

The Nuclear Data Working Group (NDWG) is comprised of nuclear data and applications experts nominated by program offices and the national laboratories and works to facilitate collaboration between multiple funding agencies on cross cutting nuclear data efforts. The primary effort of the NDWG is the organization of the annual Workshop for Applied Nuclear Data Activities (WANDA), which includes road mapping sessions on critical nuclear data topics with the goal of providing recommendations for nuclear data activities that will have a positive impact on multiple applications. The recommendations from these workshops are captured in published reports. The reports and more information can be found at the NDWG website at https://www.nndc.bnl.gov/ndwg/.

## 1.3. WoNDRAM: WORKSHOP ON NUCLEAR DATA FOR REACTOR ANTINEUTRINO MEASUREMENTS

WoNDRAM was organized by the Nuclear Data Working Group (NDWG) to bring together the nuclear data, reactor modeling, and antineutrino physics communities to address the nuclear data that reduce the uncertainties in the reactor antineutrino measurements. The workshop included three topical sessions: 1) Nuclear data for the reactor source term calculation, 2) nuclear data for the antineutrino spectrum calculation, and 3) nuclear data for detector response calculations. There were over 180 participants in attendance. Each topical session introduced the problem, and experts in the field presented the state of the art and their nuclear data needs. The agenda and presentations can be found on the NDWG workshop webpage: https://www.nndc.bnl.gov/ndwg/workshops.html. The session leads facilitated discussion amongst the participants to gain consensus on the highest priorities for nuclear data and potential solutions. The result was recommendations on nuclear data activities that will reduce uncertainties in the reactor antineutrino measurements, which are captured in the following sections of this report. Improved nuclear data may lead to improved predictions of the measured antineutrino spectrum leading to greater understanding of antineutrino physics. Additionally, improved confidence in the reactor antineutrino spectrum calculations may lead to new capabilities for reactor monitoring.

## 2. SUMMARY OF HIGHEST PRIORITY NUCLEAR DATA NEEDS

The three nuclear data road mapping sessions each examined a different part of the reactor antineutrino measurement system: 1) the reactor source term, 2) the antineutrino spectrum calculation, and 3) the detector response. When prioritizing nuclear data needs, the following goals were considered: 1) better understand the reactor anomaly and 2) improve the ability to monitor reactors using antineutrinos. How does the nuclear data improve the understanding and ability to model the antineutrino reactor anomaly, and how does the nuclear data improve the ability to apply antineutrino measurements to reactor monitoring. With these goals in mind, the highest priority needs are summarized here:

1. Perform new electron spectra measurements following the neutron-induced fission of $^{235,238}$U and $^{239,241}$Pu. For $^{235}$U and $^{239,241}$Pu, high thermal neutron fluxes from the Oak Ridge National Laboratory (ORNL) or the National Institute of Standards & Technology (NIST) research reactors would be needed. For $^{238}$U, the large, well-characterized fast neutron fluxes of the Triangle Universities Nuclear Laboratory (TUNL) particle accelerator would be needed. These measurements would take advantage of superconducting solenoids, solid-state detectors, and digital electronics–features not available in the existing measurements–to obtain total uncertainties of 1% or less in the 2 to 8 MeV electron energy range.
2. Perform correlated high-statistics antineutrino measurements at U.S.-based LEU and HEU reactors with a common high-precision detector; alternatively, perform an improved high-statistics $^{235}$U antineutrino flux and spectrum measurement at the accessible and high-power ORNL or NIST HEU research reactors.
3. An updated sensitivity study of the reactor source term to include fission yields as a function of reactor neutron spectra, fission product decay and capture cross sections, contributions from production and decay of minor actinides and capture on detector materials. This recommendation is based on the fact that the contribution of many interactions is assumed to be minimal, but there has not been a systematic study to determine the influence of many of these interactions.
4. Improve the understanding of scintillation yields and quenching factors for detector media. This can be accomplished through measurements, theory development, creation of reporting standards and creation of a library.

## 3. WoNDRAM ROAD MAPPING SESSIONS AND RECOMMENDATIONS

The session goals, prioritized needs, and recommendations are provided in more detail in this section.

### 3.1. NUCLEAR DATA NEEDS FOR REACTOR SOURCE TERM FOR REACTOR ANTINEUTRINO MEASUREMENTS

***Session Leaders:***
- Andrew Conant, ORNL
- Catherine Riddle, Idaho National Laboratory (INL)
- Will Wieselquist, ORNL

***Rapporteurs:***
- Logan Scott, ORNL
- Chad Alan Lani, The Pennsylvania State University (Penn. State)

### 3.1.1 Session Goals

The goal of the Reactor Source Term session of the Workshop on Nuclear Data for Reactor Antineutrino Measurements (WoNDRAM) is to focus nuclear data needs towards those that overlap antineutrino spectra needs with the capabilities of nuclear reactor modeling and simulation, operation, and applications. The session covered all issues related to nuclear data uncertainties up until the fission process and the effects of other nuclear reactions on the subsequent neutrino spectra; beta spectra and their subsequent conversion were covered in the next session. Examples of topics covered include fission yields, non-fuel effects, advanced reactors, and other reactor-specific effects that may alter the neutrino spectrum. Improvements in modeling simulation capabilities and integration with the beta and neutrino spectra were also discussed.

### 3.1.2 Introduction

In the Reactor Source Term session, effects up until the fission process were considered. This included the independent fission product yields and fission fragment decay. Other factors included neutron reaction cross-sections and operational considerations that impact the fission yields and thus the antineutrino spectrum. Data related to the individual decay branches for β were not considered. Time-dependent effects such as the non-equilibrium corrections were also considered. Potential applications to advanced reactor calculations and streamlining calculations with modeling and simulation tools was also desired.

### 3.1.3 Session Summary

The session covered three primary topics: 1) fission yields, 2) absorption cross section data, and 3) non-equilibrium corrections.

#### 3.1.3.1 *Fission Yields*

One of the most significant sources of uncertainty in nuclear reactors that affects the antineutrino source term is in the area of independent fission yields. In most cases, cumulative fission yields are measured, and independent yields are inferred. Currently, the JEFF-3.3 fission yield database is the preferred source of yields for antineutrino applications. Isomeric splitting is a source of uncertainty when calculating the antineutrino spectrum. However, there is an ongoing effort to measure and reevaluate the fission yields of $^{235}$U, $^{238}$U, and $^{239}$Pu as a function of neutron energy. An issue with current JEFF-3.3 neutron induced fission yield libraries is their evaluation at only three incident neutron energies–thermal, fast (400 keV), and 14 MeV. ENDF-VIII.0 has a similar three-group scheme but has a fast evaluation given at 500 keV and includes a 2 MeV evaluation for $^{239}$Pu. Yields from fissions caused by neutrons between these energy points are interpolated or assumed to be close to the yields at a nearby energy. However, for thermal reactors, the behavior of the fission yields in the epithermal/resonant region can vary and is not well known. Past work has shown as much as a twenty percent difference in the yield of important isotopes in the fission resonances.[1] Of specific note are the Pu yields at 200-300 eV neutron energy. It is not known if small differences in the reactor neutron spectrum can create measurable changes in the fission product production in reactors. Integral experiments with shifts in the reactor neutron spectrum may provide information on the effects of the neutron energy shifts. Additional energy differential measurements of fission product yields in the epithermal energy region would be beneficial for reactor models.

---

[1] Romano, Catherine E., Y. Danon, Robert C. Block, J. Thompson, Ezekiel Blain and Evelyn M. Bond. "Fission fragment mass and energy distributions as a function of incident neutron energy measured in a lead slowing-down spectrometer." Physical Review C 81 (2010): 014607.

Sensitivity studies would be useful to determine the variations in the fission yields that impact the antineutrino spectrum. To support this, fission product yield correlations to decay data with an emphasis on branching ratios would have a positive impact overall. Isomeric ratios require additional measurements to support these data and some errors in ENDF/B-VIII.0 uncertainties have been discovered which are being corrected during the new evaluation effort.

To resolve some of the issues in the fission yields, several efforts are proposed:

1. As part of the current fission yield evaluation project, provide refinement of the energy grid representation for intermediate neutron energies.

2. Independent and cumulative fission yield measurements in the epithermal range (1 eV to 100 keV). This includes energy differential and energy integral measurements.

3. Correlation studies of fission yields with decay data to support studies of uncertainties in cumulative and independent yields.

### 3.1.3.2 *Absorption Cross-Section Data*

Antineutrino yields and spectra are dependent on the fission rate and parent nuclide concentrations in a nuclear reactor. The reactor design and neutron spectrum have an impact on the contribution of different nuclides to the fission rates, which also vary spatially and temporally in the reactor. Because a significant fraction of neutrons produced in a reactor will be lost due to absorption, significant neutron absorbers alter the spectrum. Many fission product absorptions are frequently calculated and well known to the reactor physics communities (e.g., Xe capture), while notable fertile isotope capture reactions (e.g., U capture and its decay to Pu), do not produce antineutrinos above the inverse beta decay (IBD) threshold. However, low energy antineutrinos will impact coherent elastic neutrino nucleus scattering (CEvNS) measurements because they have a lower energy threshold by orders of magnitude when compared to IBD.

As an example, Rb-92 is a known fission yield whose absorption cross section affects the cumulative yields. An uncertainty study is required to identify the fission yields with absorption cross sections of concern to reactor antineutrino measurements. Incorporating the antineutrino spectrum into depletion codes such as ORIGEN will support these types of uncertainty studies.[2]

To resolve some of the issues related to absorption cross-sections, multiple efforts are proposed:

1. Sensitivity studies to identify the fission yields with absorption cross sections of concern to reactor antineutrino measurements.

2. Targeted capture cross section and decay measurements of high impact isotopes.

---

[2] SCALE Code System, ORNL/TM-2005/39, Version 6.2.4 (April 2020)

3.1.3.3 *Status of Non-Equilibrium Corrections*

The basis for the most commonly used experimental antineutrino spectra is a series of measurements performed at the Institut Laue-Langevin (ILL) in the 1980s with measurements performed on time scales of hours to days.

Fission fraction uncertainties for the Double Chooz experiment[3] were bracketed by adjusting simulation parameters using the MURE and DRAGON code.[4] These corrections can also be bracketed by comparing to post-burnup fuel content. In spent fuel safeguards studies, the amount of Pu produced in a reactor is influenced by moderator density, boron concentration, and the buildup of neutron absorbers. There are also special variations in Pu production due to control rods and structural materials within the core. An important factor that influences uncertainties in modeling is access to accurate core loading patterns, reloading parameters and boron letdown.

Reactor benchmarks can be helpful to validate the reactor codes. These data sets may be able to be stored and maintained by the International Atomic Energy Agency (IAEA). One such database is the OECD/NEA Spent Fuel Composition Database[5] (SFCOMPO) which provides spent fuel isotopic data along with reactor operating parameters. Shared sets of antineutrino measurement data and calculated reactor source term data would be useful for the community. The MURE and DRAGON calculations can be compared to the Takahama benchmark. The uncertainties in the ability to determine the fission fraction of U and Pu are required to inform future work.

To resolve some of the issues related to non-equilibrium effects, several efforts are proposed:

1. Sensitivity studies to calculate isotopes that impact to the non-equilibrium antineutrino spectrum.
2. Integration of the neutrino spectrum within depletion codes to support studies of reactor operations and antineutrino spectra.
3. Create a database of reactor operating parameters, simulations, and measurements relevant to the antineutrino community.

### 3.1.4 Summary of Recommendations

In order to improve the calculation of the reactor antineutrino source term which support antineutrino measurements, the following efforts are proposed:

1. Perform measurements of cumulative fission yields in the epithermal range to understand the variation of fission yields in different resonance regions. This can be accomplished with energy differential measurements in a facility such as the lead slowing down spectrometer at RPI, which can provide good statistics, or the SPIDER detector at Los Alamos National Laboratory (LANL).

---

[3] http://doublechooz.in2p3.fr/Scientific/scientific.php
[4] C. L. Jones, A. Bernstein, J. M. Conrad, Z. Djurcic, M. Fallot, L. Giot, G. Keefer, A. Onillon, and L. Winslow, "Reactor simulation for antineutrino experiments using DRAGON and MURE", Phys. Rev. D 86, 012001– Published 10 July 2012. arXiv:1109.5379v4.
[5] Germina Ilas, Ian Gauld, Pedro Ortego and Shuichi Tsuda, "SFCOMPO DATABASE OF SPENT NUCLEAR FUEL ASSAY DATA–THE NEXT FRONTIER", EPJ Web Conf., 247 (2021) 10019.
DOI: https://doi.org/10.1051/epjconf/202124710019

Energy integral measurements can be performed at reactors with modified spectra. The priority isotopes are $^{235}$U, $^{238}$U, $^{239}$Pu, and $^{241}$Pu.

2. Refine the energy grid representation for evaluated FPY data at different incident neutron energies.

3. Continue fission yield measurements to include several methods that provide independent yield, cumulative yields, neutron energy differential, and integral data.

4. Support sensitivity studies on the impact of small neutron spectrum shifts on the changes in resonance yields.

5. Support sensitivity studies to understand the largest contributions to the non-equilibrium effects. The non-equilibrium effects, which account for fuel evolution, need to be evaluated on a realistic neutron energy distribution.

6. Support sensitivity studies to understand the influence of neutron capture cross sections on the aggregate antineutrino spectrum. This would also be useful for code enhancements to the reactor community.

7. Establish an independent database for the sharing of reactor parameters and associated antineutrino spectra. Major parameters with effects on the antineutrino spectrum which could be included are moderator type and density, fuel type and operating temperature, and geometry.

## 3.2. NUCLEAR DATA AND ANTINEUTRINO SPECTRA SESSION

*Session Leaders*:
- Patrick Huber, Virginia Tech (VT)
- Bryce Littlejohn, Illinois Institute of Technology (IIT)
- Shikha Prasad, Texas A&M University (TAMU)
- Alejandro Sonzogni, Brookhaven National Laboratory (BNL)

*Rapporteurs*:
- Anosh Irani, IIT
- Wei-Eng Ang, TAMU

### 3.2.1 Session Goals

Antineutrinos produced by nuclear reactors are useful for non-proliferation and reactor monitoring applications, for validating other existing and future datasets in the nuclear data pipeline, and for performing fundamental particle physics measurements. These three spheres of application rely to varying degrees on an accurate and precise understanding of the true aggregate antineutrino energy spectrum emitted by the daughters of each fission isotope. The goals of this session were to identify future experimental, theory, and software improvements that can expand understanding of directly measured and indirectly predicted antineutrino spectra per fission, and to define the extent to which each of these improvements will benefit the three spheres of application described above.

### 3.2.2 Introduction

The WoNDRAM 2021 Session on Nuclear Data and Antineutrino Spectra focused on identifying future experimental, theory, and software improvements that can expand understanding of directly measured and indirectly predicted antineutrino spectra emitted per fission from conventional, research, and advanced reactors. Effort was also made to connect specific improvements to attendant benefits in fundamental physics, nuclear data validation, and applied antineutrino technology use cases.

This session was organized into four main segments. The first segment was intended to summarize existing methods for predicting antineutrino spectra for each fission isotope and to assess the value of model improvements for the three use cases. The following three segments were dedicated to highlighting methods or measurements that could improve each of three antineutrino spectrum prediction methods discussed: the conversion approach, the summation approach, and the direct measurements approach. The user briefs identified a broad set of needs for these three differing approaches for estimating antineutrino spectra emitted emissions from nuclear reactors. These are addressed in turn below, in addition to an overview of the introductory sub-session and identification of synergies.

### 3.2.3 Uses Cases–"Where Are the Needs?"

A series of introductory briefs on Day 1 and Day 3 established that the nuclear data validation, applied antineutrino technology, and fundamental physics use cases substantially vary in their reliance on future improvements in the understanding of the antineutrino release per fission for different isotopes. For the fundamental physics use case, specialized experimental arrangements have enabled measurements of neutrino properties at reactors despite the lack of precise understanding of the relevant source term, and as such, these and foreseeable future measurements do not depend much on improvements in this area. Conversely, the nuclear data validation use case has expanded tremendously in scope largely due to vast improvements in direct reactor antineutrino measurement precision in the past decade and will likely continue its expansion with attendant improvements in both nuclear and neutrino datasets. As the presented briefs illustrated, the applied antineutrino technology use case has relied largely on theoretical predictions to demonstrate potential capabilities. However, realized applications require real-world demonstrations, particularly when predicted capabilities rely on demonstrably imperfect models. Given the high-stakes nature of reactor monitoring in a treaty context, an exceptionally high degree of reliability is required. In particular, any neutrino flux model will need to be fully validated by direct neutrino measurements. Apart from reliability, more precise capabilities in predicting antineutrino emission from individual fission isotopes (including higher actinides) and reactor types (such as advanced reactors and accelerator-driven systems) would enhance the quantitative conclusions that can be drawn in reactor monitoring. In this sense, this use case benefits from improvements in both models and direct measurements.

*3.2.3.1 Conversion Model Predictions*

The conversion prediction method iteratively fits individual ad hoc beta decay spectra (so-called "virtual branches") to a measurement of the aggregate fission beta spectrum for each fission isotope and then converts the ad hoc spectra into an antineutrino energy spectrum per fission using energy conservation and including higher-order corrections. Current conversion predictions can be improved by enhancing the quality of input aggregate fission beta spectrum data or by honing or validating theoretical assumptions regarding the shape of fitted virtual beta spectra.

The 1980s ILL-measured fission beta spectrum measurements underpinning current conversions have received intense recent criticism from both the nuclear physics and neutrino physics communities. ILL-based conversions incorrectly predict observed changes in neutrino emission rates with LEU fuel burn-up, while summation models predict this behavior correctly. In addition, a recent low-precision Russian measurement of fission beta yield ratios between $^{235}$U and $^{239}$Pu was found to be inconsistent with ILL results. In light of these developments, there was broad general agreement among attendees that new, modern, high-statistics measurements of fission beta spectra would be particularly valuable. A series of user briefs discussed the experimental methods required to feasibly perform such a measurement at a U.S.-based neutron facility and the extent to which these measurements would exceed ILL's in reliability, precision, and breadth. *Performance of new high-precision aggregate fission beta spectrum measurements for many isotopes is thus highly recommended (recommendation 1).*

Fitted beta spectrum shapes used in conversion predictions rely on a variety of theoretical or database-driven inputs. Of these inputs, briefs in this session identified one particularly high-priority item: shape factors for forbidden decays, which contribute much of the reactor antineutrino spectrum at high energies. *Thus, it was also recommended that direct measurement of beta spectrum shapes be performed for selected forbidden transitions, particularly ones generating outsized contributions to the high-energy reactor antineutrino spectrum (recommendation 2).*

### 3.2.3.2 Summation Model Predictions

The summation method produces a predicted antineutrino spectrum per fission for each fission isotope by summing contributions from each individual beta decay branch of each fission product using fission yield nuclear databases (such as JEFF 3.3) and beta decay databases (such as ENDF B-VIII.0). Improvements in the breadth, quality, and precision of data in these evaluated databases thus leads to enhanced accuracy in summation-predicted antineutrino spectra. Recent and continuing beta decay measurement campaigns using Total Absorption Gamma Spectroscopy have already substantially improved beta feeding data for many of the isotopes prominently contributing to the problematic 5-7 MeV range of the antineutrino spectrum, while development of an improved ENDF fission yield evaluation is underway. Both fission yield and beta feeding project spheres currently benefit from NDIAWG-related support obtained in previous funding cycles.

Despite recent improvements, challenges remain. Reliable error budgets for summation model predictions remain elusive and the comparison with the aggregate beta spectra or directly measured neutrino spectra instead is used as a proxy. Continued advancement in the correction of pandemonium-affected nuclear data is needed for accurate summation predictions of reactor antineutrino spectra at the highest (~7-12 MeV) energies and for accurate prediction of fine spectral structure in future high-resolution antineutrino measurements. Finally, important aspects of fission yields and beta decays relevant to the antineutrino spectrum remain critically under-studied, such as isomeric yield ratios for many high-Q isotopes and decay shape factors for high-Q forbidden transitions. To enable continued progress, support for respective projects should be continued. To enable future improvements, *highest recommendations include the performance of direct measurements of beta spectrum shapes for selected forbidden transitions (recommendation 3).*

### 3.2.3.3 Direct Source Term Measurements

Both summation and conversion models predict antineutrino detection rates and energy spectra at odds with directly measured reactor antineutrino datasets. Measurements at both HEU and LEU reactors have observed IBD detection rate deficits with respect to models and large deviations from model-predicted IBD spectrum shapes. Improvement and standardization of directly measured antineutrino datasets can help to further elucidate the nature of modelling inaccuracies and can enable precise and unbiased reactor antineutrino source term predictions for widely varying reactor types.

While the precision of IBD-based measurements of antineutrino production for LEU reactors already exceeds that of summation and conversion models above the 1.8 MeV IBD threshold, IBD-based measurements of antineutrino production by individual fission isotopes remain comparatively imprecise and must be improved. Direct measurement precision for the dominant isotopes $^{235}$U and $^{239}$Pu can be enhanced with improved short-baseline antineutrino measurements at an HEU and an LEU reactor, respectively. Improved direct $^{238}$U constraints can be achieved through combined analysis of these measurements, particularly if both are performed using the same detector. Precise antineutrino measurements at future advanced reactors may enable meaningful direct constraints for other sub-dominant isotopes, such as $^{241}$Pu and $^{240}$Pu, as well as for fast fissions. These new measurements provide yet further value for elucidating the quality of summation predictions if energy resolution of a few percent or better can be achieved. To enable future pursuit of direct constraints on reactor antineutrino production below the 1.8 MeV IBD threshold, further research and development (R&D) on reactor-based coherent neutrino scattering (CEvNS) and electron scattering detectors must be performed. *Thus, performance of improved new high-precision antineutrino spectrum and flux measurements at HEU and LEU reactors, preferably in a systematically correlated manner, was highly recommended (recommendation 2 and 6). It was also recommended to support existing R&D efforts dedicated to detecting reactor antineutrinos with threshold-free interaction channels, such as CEvNS or electron scattering (recommendation 4).*

Session briefs also identified technical hurdles to realizing the full utility of existing and future direct antineutrino measurements for nuclear data and reactor applications, including the difficulty of assessing consistency between IBD-based datasets and between IBD datasets and updated models, and the lack of public availability of these datasets. *Thus, recommendations also included establishing standardized evaluations of reactor antineutrino datasets and predictions in a centralized, publicly accessible domain (recommendation 5).*

### 3.2.4 Synergies

Since it remains difficult to provide an error budget for summation calculations the comparison with aggregate beta spectra is an important tool to assess the validity and completeness of any summation calculation. For conversion calculations, the aggregate beta spectra are the central ingredient. Therefore, improved aggregate beta spectra benefit both approaches for flux prediction. As far as the 5 MeV bump is concerned, a remarkable observation is that both summation and conversion models agree on the size, shape, and position of the bump. Therefore, any direct neutrino measurement providing more information about the bump and its fissile isotope dependence, if it exists, would be valuable for both approaches to flux predictions. The following recommendations would specifically support nuclear data and reactor monitoring end users. *Their value in improving all three methods of understanding fission antineutrino emissions further strengthens the case for highly recommending the new high-precision fission beta and antineutrino measurements (recommendations 1, 2, and 6).*

The development of an open software framework enabling uncertainty quantification for and standardized comparisons between direct neutrino measurements and conversion and summation calculations would greatly facilitate progress and reduce hurdles of participation for use case communities. The following recommendations would benefit all three discussed end user communities: nuclear data, reactor monitoring, and particle physics. *That broadened access and utility can be delivered to both the predictions and direct measurement communities by these tools further strengthens the recommendation for their development (recommendation 5).*

A major limitation for both conversion and summation calculations is our lack of understanding of forbidden decays and their associated shape factors, since as much as 40% of neutrinos arise from forbidden decays. First attempts to use theoretically computed shape factors have been made and the resulting corrections are at the percent-level. Validating these theoretical shape factors for relevant fission fragment isotopes is a crucial step to enhance the reliability of both conversion and summation calculations. The following recommendations would directly support nuclear data end users and would directly benefit reactor monitoring end users. *Their value in improving both conversion and summation predictions strengthens the case for recommending the beta shape factor measurements (recommendation 3).*

The main stumbling blocks towards a reliable conversion flux calculation are well-founded doubts about the validity of the aggregate beta spectra measured in the 1980s at the ILL and our lack of a quantitative understanding of the impact of forbidden decays. At the same time, none of our predictive tools can reproduce the 5 MeV bump: only direct neutrino source term measurements can accurately address this phenomenon. Summation calculations offer a valuable tool to address higher order corrections like non-equilibrium effects or the contribution of structural materials in the reactor core, but at this time, the calculations lack error estimates. A comparison of summation calculations with both direct source term data and aggregate beta spectra would serve to establish the validity of the method and employed data. Therefore, only a combination of high-precision aggregate beta spectrum measurements, antineutrino measurements, and beta shape factor measurements for relevant isotopes can provide sufficient information to assure that the nuclear data and reactor monitoring end-uses are fully supported. Leaving out any one of the three proposed components above substantially reduces the value obtained from the remaining components.

### 3.2.5  Consensus Priorities for Future Work

The utility of reactor antineutrino emissions for nuclear data validation, non-proliferation and reactor monitoring applications, and particle physics use cases is dependent on an accurate and precise understanding of reactor neutronics and content, of antineutrino production per fission, and of antineutrino detector response–the primary topics addressed in each of the three WoNDRAM discussion sessions. Recommendations for improving understanding for each of these categories were presented in the previous section.

Inaccuracies in modelled reactor antineutrino detection signatures have been clearly established through comparison of existing measurements and model predictions at HEU and LEU cores. Specifically, models fail to reproduce energy spectra and fluxes matching those measured in IBD-based detectors. Of the three model input categories described above, incorrect understanding of neutrino production per fission is most likely the primary cause of these discrepancies. *For this reason, recommendations related to improving understanding of antineutrino production per fission deserve special attention.*

### 3.2.6  Summary of Recommendations

To resolve current disagreements between reactor antineutrino measurements and predictions, the following recommendations are made. The value of these recommended items is maximized if most, or all, are undertaken and successfully achieved:

1. Perform new electron spectra measurements following the neutron-induced fission of $^{235,238}$U and $^{239,241}$Pu. For $^{235}$U and $^{239,241}$Pu, high thermal neutron fluxes from the ORNL or NIST research reactors would be needed. For $^{238}$U, the large, well-characterized fast neutron fluxes of the TUNL particle accelerator would be needed. These measurements would take advantage of superconducting solenoids, solid-state detectors, and digital electronics–features not available in the existing measurements–to obtain total uncertainties of 1% or less in the 2 to 8 MeV electron energy range.
2. Perform correlated high-statistics antineutrino measurements at U.S.-based LEU and HEU reactors with a common high-precision detector; alternatively, perform an improved high-statistics $^{235}$U antineutrino flux and spectrum measurement at the accessible and high-power ORNL or NIST HEU research reactors.
3. Perform electron spectra measurements for the largest fission product contributors to the total antineutrino spectrum, to quantify the effect of sub-dominant corrections to the spectrum shape. The required precision would be 2% or better. These measurements would take place in the CARIBU facility at ANL for fission products such as $^{92}$Rb, $^{96}$Y, and $^{100}$Nb, which are the largest contributors according to the nuclear databases. Precise β-energy shape measurements of individual β-feedings to individual energy levels for isotopes with high Q-value and with high cumulative fission yields have never been published.

The following recommendations will also enhance prospects for understanding and utilizing fission-produced reactor antineutrinos:

4. Support existing R&D efforts dedicated to detecting reactor antineutrinos with threshold-free interaction channels, such as CEvNS or electron scattering.
5. Establish standardized evaluations of reactor antineutrino datasets and predictions in a centralized, publicly accessible domain. Currently, the nuclear databases are stored in the legacy ENDF/B format, while the measured reactor antineutrino spectra are not stored anywhere; additionally, the codes to calculate the electron and antineutrino spectra following decay are numerically heavy and are not publicly available.
6. Support dedicated joint analysis of existing antineutrino datasets from HEU and LEU reactors. For instance, a recent analysis of the Daya Bay and PROSPECT data resulted in considerably more precise $^{235}$U and $^{239}$Pu IBD antineutrino spectra.

## 3.3. NUCLEAR DATA NEEDS FOR DETECTOR RESPONSE FOR REACTOR ANTINEUTRINO MEASUREMENTS

*Session Leaders:*
- Nathaniel Bowden, Lawrence Livermore National Laboratory (LLNL)
- Bethany Goldblum, Lawrence Berkeley National Laboratory (LBNL)/UC Berkeley
- Jon Link, VT
- Pieter Mumm, NIST

*Rapporteurs:*
- Karla Tellez-Giron-Flores, Virginia Tech
- Eric Matthews, UC Berkeley

### 3.3.1 Session Goal

The goal of this session was to identify and prioritize nuclear data needs that impact antineutrino detector modeling capabilities in the reactor energy range–important for detector design, development, and data interpretation. Discussion highlighted nuclear data deficiencies in modeling both the neutrino signal and background for coherent elastic neutrino nucleus scattering, inverse beta decay, and electron-neutrino scattering measurement approaches.

### 3.3.2 Introduction

Advances in detector modeling capabilities would further the development of basic and applied antineutrino physics technologies. This session focused on identifying and prioritizing nuclear data needs that impact the ability to model antineutrino detector performance in the reactor energy range–important for detector design, development, and data interpretation. In this context, nuclear data encompasses "traditional" nuclear data that underlie modeling and analysis (e.g., cross sections, level schemes, etc.), "non-traditional" nuclear data (e.g., scintillator response, quenching factors, etc.), improved tools and models needed for reactor antineutrino detector response, and compilations and evaluations of data relevant to the above.

The session was organized into two main segments: nuclear data needs specific to different measurement approaches and crosscutting nuclear data needs. The coherent elastic neutrino nucleus scattering, inverse beta decay, and electron elastic scattering detection techniques were examined, and data deficiencies were reported in the context of modeling both neutrino signal and background.

### 3.3.3 Session Scope and Identified Needs

The user briefs identified a wide range of data needs. The needs and issues are summarized below according to their subject area within the broader scope of reactor antineutrino measurements.

*3.3.3.1 Coherent Elastic Neutrino-Nucleus Scattering*

While coherent elastic neutrino-nucleus scattering (CEvNS) was proposed by Freedman in 1974,[6] its observation in recent years[7] has led to an increased interest in CEvNS-based reactor antineutrino measurements.[8] Quenching factors for scintillators commonly used in CEvNS measurements were cited as a high priority need in this regard. In addition to being required input for the accurate modeling of scintillator-based detection systems, knowledge of material quenching is critical for the interpretation of CEvNS neutrino studies. For example, recent work by Khan and Rodejohann demonstrated that improved quenching factor knowledge applied to COHERENT's measurement of CEvNS could increase the physics reach of the experiment.[9] However, the quenching data used in that work has been questioned,[10] highlighting the need for reliable quenching measurement techniques to aid experimental interpretation. In addition to measurement inconsistencies, significant discrepancies have been observed between models and experimental data showcasing the need for improved theoretical modeling of detector material quenching. The compilation of nuclear recoil quenching factors in the form of a database was cited as a potential means to address inconsistencies in experimental data through evaluation. Likewise, the need for dedicated experimental capabilities and additional beam time at accelerator facilities was noted.

---

[6] D.Z. Freedman, Phys. Rev. D 9, 1389 (1974).
[7] D. Akimov, et al., Science 357, 1123 (2017).
[8] M. Bowen and P. Huber, Phys. Rev. D 102, 053008 (2020).
[9] A.N. Khan and W. Rodejohann, Phys. Rev. D 100, 113003 (2019).
[10] D. Persey, *"New CEvNS Results from the COHERENT CsI[Na] Detector"*, https://theory.fnal.gov/events/event/new-results-from-coherent-2/

The contributors to background in CEvNS measurements include cosmogenics, site-related radiation fields, ambient and intrinsic radioactivity, and detector-specific noise and dark rate. Neutrons are particularly problematic in that fast neutron-induced nuclear recoils are indistinguishable from the CEvNS signal, and neutron fields tend to be very dependent on detectors, sites and atmospheric conditions, and shielding configurations. Furthermore, neutron backgrounds can be correlated with the source (reactor thermal power in the case of reactors, in time for spallation sources). Cosmogenic muon-induced neutrons produced in detector shielding and surrounding structures are a significant source of background and in many cases, must be predicted using simulation[11]. As will be noted for IBD, cosmogenic neutron simulations rarely agree with data and ad-hoc scaling is often required. A novel background for CEvNS, albeit likely to be subdominant, is inelastic neutrino nucleus interactions. These reactions are important for CEvNS measurements with reactor neutrinos as neutrino-induced neutrons and de-excitation gamma-rays from inelastically excited nuclei may represent unknown background contributors. There are large variations in theoretical predictions of the inelastic neutrino nucleus cross section and direct measurements for reactor energy neutrinos are difficult.

*3.3.3.2 Inverse Beta Decay*

The inverse beta decay (IBD) interaction wherein an electron-flavor antineutrino interacts with a proton to produce a positron and a neutron has features that strongly motivate its use in reactor-based neutrino experiments. The IBD cross section is well understood and rises quickly over the fission energy range. Neutrino energy information is carried by the positron and the correlated neutron provides an event tag critical to efficient event selection and background rejection. Technologies have progressed significantly over the past decades[12] yet several nuclear data needs were identified as required to support further progress.

*3.3.3.3 IBD Signal*

As already noted, the IBD cross section on free protons in organic target materials is well known, and therefore this aspect of detector design does not represent a data need. The data needed to predict the manifestation of an IBD signal event in a detector is generally thought to be adequate for system design purposes, and although improvements are possible, they are not considered to be a high priority. As noted in the "Cross-cutting Data Needs" section (section 3.3.4) of this report, knowledge of quenching factors for the prompt positron (electromagnetic) deposition influences the ability to reconstruct incident neutrino energy. Estimation of IBD event selection efficiency depends upon 1) neutron transport modeling and 2) detector modeling of final state particles from neutron capture processes (e.g., $^6$Li (n,$\alpha$)t, Gd (n,n'$\gamma$) Gd*). The former case is discussed elsewhere in this section, as are quenching factors relevant to capture reactions with heavy ion products. The capture cross section for these processes is well known. Existing statistical models of the **γ**-ray cascade following neutron capture on Gd are thought to be sufficient for design purposes[13] for both organic scintillator and water-based detection media.

---

[11] Hakenmüller, J., et al. Neutron-induced background in the CONUS experiment. Eur. Phys. J. C 79, 699 (2019). https://doi.org/10.1140/epjc/s10052-019-7160-2

[12] A. Bernstein, N. Bowden, B.L. Goldblum, P. Huber, I. Jovanovic, and J. Mattingly, "Colloquium: Neutrino Detectors as Tools for Nuclear Security," Rev. Mod. Phys. **92**, 011003 (2020).

*3.3.3.4 Aboveground Background*

Recent questions of fundamental neutrino physics and reactor monitoring have motivated short baseline measurements at or near the Earth's surface. The deployment of neutrino detectors with minimal overburden poses special challenges, and aboveground detection of reactor antineutrinos has been demonstrated only recently. At-surface background interaction rates are relatively high, and there exist multiple physical processes that can mimic an IBD signal. The primary concern for achieving a sufficient signal-to-background rate for a useful detector is supporting a sufficiently detailed and accurate model of both the processes themselves and of the detector response such that differences can be exploited as analysis constraints with well-determined efficiencies. *An improved understanding of scintillator response, including quenching models, stopping powers for heavier nuclei, and scintillation yields were identified as essential data needs in this regard.* Presently, no quenching database exists, thus each experiment must build and validate its own quenching model. Measurement approaches and data formatting are not standardized, which further reduces research efficiency and has led to inconsistent results.

When backgrounds cannot be directly measured (e.g., during reactor off periods), source terms of sufficient accuracy may be used in a direct subtraction. To facilitate background subtraction in such situations, there is the need to accurately model background progenitors, often cosmogenic in origin, in potentially complex environments and transport them throughout the detector environment. In the case of surface deployment, the majority of IBD-like backgrounds are the result of fast neutron interactions in the detector. Neutrons can yield a variety of event classes (e.g., multiple correlated neutron captures [vetoed by identifying recoils and possibly topology]), inelastic scatter just above threshold where the associated nuclear recoil is too small to be seen in the detector, and incorrectly tagged recoils followed by capture. Rates depend on accurately understanding the source term. In the case of muon-induced neutrons (e.g., correlated multi-neutron shower production by muon spallation), neutron yields are currently not well modeled. *Improved neutron yield data were identified as a need.* The need to include multi-GeV hadronic interactions and showers was also noted. In addition to the source terms, a full modeling of neutron transport over many orders of magnitude in energy is required. Some measurements require a detailed understanding of thermalization (e.g., cross section resonances at low energy), efficiencies that depend on capture time or distance, and care must be taken with the scattering cross-sections used. Molecular effects can become important and are not always available at the precision needed. In terms of neutron yields at higher energies, cross sections above 20 MeV are mostly unmeasured, and codes must depend on models. Improved measurements of these processes would have a positive impact on predictive background modeling for surface-based neutrino detectors.

To have the greatest utility in background subtraction and experimental design, the results of background modelling must also have associated uncertainties. There are no methods for propagation of nuclear data uncertainties presently integrated in Monte Carlo codes like Geant4. *Incorporation of nuclear data*

---

[13] e.g., H. Almazan, et al (STEREO) "Improved STEREO simulation with a new gamma ray spectrum of excited gadolinium isotopes using FIFRELIN", Eur. Phys. J. A (2019) 55: 183, http://dx.doi.org/10.1140/epja/i2019-12886-y

*uncertainties in simulation libraries and uncertainty propagation methods in simulation frameworks was identified as a cross-cutting development need for all types of neutrino detectors.*

*3.3.3.5 Belowground Background*

As with the aboveground case, the primary concern when designing an experiment is background estimation. Uncertainties in background estimates can necessitate a conservative design approach, potentially increasing the cost and construction timeline. There are three major categories of background that are relevant. First, accidental coincidences from intrinsic radioactivity can be controlled by careful material selection, cleanliness procedures during construction, and analysis techniques. Second, cosmogenic production of beta-neutron emitting radionuclides can provide antineutrino mimicking events. This class of event can often be tagged by detecting the initiating muon. Third, cosmogenic production of fast neutrons in material outside a detector can be problematic if that neutron "punches through" into the active volume generating a recoil signal or inelastic gamma-ray followed by a neutron capture.

Experimental design of belowground detectors is complicated by the variation of muon energy and rate with depth and the poorly characterized neutron and radionuclide yields as a function of these parameters. For example, Geant4 and FLUKA do not agree in predicting neutron yields.[14] There are a limited number of measurements at different depths, and these are often from very different systems. Background variation and response may also be different for relatively new detection media like water-based liquid scintillator (WbLS) and Gd-doped materials. *Two classes of dedicated measurement were identified to improve this situation: fast neutron yields at different depths and radionuclide production measurements with different media using a muon beam.*

It was also noted that there may be a cost advantage to performing such measurements to inform the design of a future large below ground detector. However, the data needed in this case is sufficiently specific such that the effort should likely be included as part of that experimental program. That is, there is not a broad, high priority data need associated with belowground antineutrino detectors for application measurements.

*3.3.3.6 Neutrino-Electron Elastic Scattering*

Neutrino-electron elastic scattering is an interaction used for a variety of fundamental neutrino physics studies. There is some interest in its use for reactor neutrino detection since the scattered electron direction is highly correlated with the incident neutrino direction.[15] However, this would be an exceptionally challenging task and there are no active efforts proposing to develop a neutrino-electron elastic scattering detector for reactor neutrino detection. In this context, current capabilities to predict signal and background are considered adequate by workshop participants and *there are no pressing nuclear data needs*.

The need for nuclear data in this context relates to signal and background estimation in service of detector design activities. With respect to signal estimation, the cross section for the neutrino-electron elastic scattering interaction is calculable for relevant media to a fraction of a percent. The key technical challenges for this approach in the reactor energy range is background reduction since any electron-like energy deposition resembles the final state of elastic scatter interaction.

---

[14] F. Sutanto, et al, "Measurement of muon-induced high-energy neutrons from rock in an underground Gd-doped water detector", Phys. Rev. C 102 (2020) 034616, http://dx.doi.org/10.1103/PhysRevC.102.034616

[15] D. Hellfeld, et al, "Reconstructing the direction of reactor antineutrinos via electron scattering in Gd-doped water Cherenkov detectors" Nucl. Inst. Meth. A 841 (2017) 130; https://doi.org/10.1016/j.nima.2016.10.027

Backgrounds can be generated or emitted by the radioactive decay of detector components. The nuclear data describing backgrounds of this type are well known and do not present a data need. Instead, careful material selection and assay, strict cleanliness procedures, and unprecedented control of radon are where effort would need to be invested. Another source of background is cosmogenic isotope production by muon spallation on $^{16}$O in water and $^{12}$C in organic materials. FLUKA simulations show good agreement with data in predicting the characteristics of this mechanism, and Super-K has directly measured the total yield of cosmogenic isotopes to the few percent level. Furthermore, improved data can be expected from Super-K-Gd. In combination, these provide a sound means for experiment design and therefore estimation of cosmogenic isotope production also does not represent a nuclear data need for neutrino-electron elastic scattering.

### 3.3.4  Cross-cutting Data Needs

An understanding of scintillator light yield is required to generate realistic scintillation detector response functions. In addition to the quenching factor needs cited in the context of CEvNS and IBD detection techniques, the need for an improved understanding of ionization quenching in scintillators was presented, emphasizing the necessity for additional measurements and measurement standards and accurate and predictive modeling. The lack of core support for scintillator response studies at universities and national labs was noted, which has left the responsibility for measurements to specific projects that require these data for detector response modeling. Inconsistent treatment of electron light nonproportionality[16] was cited as a significant source of bias in the interpretation of proton quenching factors. Deficiencies were also identified regarding the physical modeling of scintillator response. The value of a database of energy-differential scintillator response measurements for both commercial and novel materials was highlighted in this regard.

The need for improved stopping power measurements in scintillators was further cited given that semi-empirical and physics-based models of scintillator quenching rely upon knowledge of the stopping power for accurate interpretation. For example, only one measurement of the stopping power of protons in the region of the Bragg peak exists for polyvinyl toluene, a common base for plastic scintillators, and no data are available for carbon ions.[17] This was noted as cross-cutting data needed for National Aeronautics and Space Administration (NASA) with regard to planetary spectroscopy. The IAEA, who bears responsibility for compilation of experimental stopping power data, urged the community to develop a prioritized list of stopping power needs to facilitate direction of future efforts.

Beyond scintillation, radiation transport calculations are also required for antineutrino detector response modeling. Deficiencies were cited for nuclear data inputs to transport codes, especially for common detector materials. This included needs for improved neutron-induced reaction cross section data on H, C, O, F, N, S, Zn, Li, B, Gd, and Cd for neutron energies greater than 20 MeV. These data were noted as crosscutting with space applications, including space-based nuclear energy and nuclear detonation detection via space platforms. In addition, the measurement and simulation of cosmic ray effects was cited as important for background characterization. Finally, the requirement for comprehensive uncertainty quantification for Monte Carlo simulation outputs, including measurement uncertainty and covariance, was noted along with options for the development of a benchmark experiment, akin to the critical assemblies for fundamental nuclear data evaluation, to enable code and library validation specific for antineutrino detection systems. Such an approach could provide information on software limitations and differences in available nuclear data cross section libraries, illuminating the need for new cross sections and/or new modeling capabilities.

---

[16] S.A. Payne, et al., IEEE Trans. Nucl. Sci. 58, 3392 (2011).
[17] C.A. Sautter and E.J. Zimmerman, Phys. Rev. 140.2A, (1965).

### 3.3.5 Summary of Recommendations

This section provides a list of recommendations followed by discussion:

1. Validation of neutron production and interaction libraries in simulation codes.
2. Incorporation of nuclear data uncertainties and uncertainty propagation in simulation codes.
3. Template of measurement standards and uncertainties for material quenching.
4. Additional measurements of electron and proton scintillation yields and quenching factors reported in conformation to the template.
5. Library of experimental scintillation yields and quenching factors.
6. Theoretical modeling of scintillator quenching.
7. Stopping power measurements for recoil particles in scintillating media, particularly heavy nuclei.

The primary challenge for reactor neutrino detection is control of backgrounds. Background modelling is essential for detector design and data interpretation. For most classes of reactor neutrino detectors, cosmogenic backgrounds are dominant. However, data-to-simulation comparisons often reveal substantial disagreement across several background classes like muon-induced neutrons, cosmogenic radioisotope production, and neutron spallation in detector materials. To address this need, it is recommended that funding agencies prioritize an effort to generate and validate neutron physics libraries. Validation may be able to leverage data from existing neutrino detectors or may require new measurements. Similarly, design and interpretation are hindered by the lack of methods for propagation of nuclear data uncertainties in Monte Carlo codes like Geant4. It is recommended that funding agencies prioritize an effort to incorporate nuclear data uncertainties in neutron physics simulation libraries and develop and incorporate uncertainty propagation methods in common simulation frameworks (e.g., Geant4).

There is a significant need for improved understanding of electron and proton scintillation yields and quenching factors. As there is currently no theoretical model to predict ionization quenching effects, scintillator response measurements are required for each scintillating medium of interest to inform detector response modeling. These data have historically fallen outside traditional concepts of nuclear data. As such, the study of scintillator response is often approached on a per project basis resulting in significant financial and time investments in the establishment of measurement capabilities as well as inconsistent analytic methods and results. This gap was previously recognized by the community as indicated in the 2020 Workshop on Applied Nuclear Data Activities (WANDA) report citing the crosscutting impacts of scintillator response for basic nuclear physics, proliferation detection, and nuclear medicine.[18]

To address this need, it is recommended that funding agencies prioritize experimental studies of scintillator response–including light yield, electron light nonproportionality, and quenching factors for heavy nuclei–as a key component of the science, as opposed to just a tool for fundamental physics inquiry. To maximize the value of these data, scintillator response measurements should be reported in conformity to a template that includes important experimental details, analytical methods, and measurement uncertainties. To facilitate this, the generation of a measurement template is urgently recommended for future scintillator response experiments. This template, similar to those produced for neutron-induced cross sections,[19] fission yields,[20] and other nuclear data observables,[21] is intended to serve as a reporting standard with expected measurement uncertainties to improve the consistency and quality of the generated data. A database compiling experimental measurements of scintillation yields and quenching factors is further recommended (akin to the EXFOR[22] nuclear reactions library). While such a library will benefit from the measurement template and experimental standards, it is recommended that compilation of scintillator response data proceed immediately as this lays the foundation for data evaluation and modeling efforts. Relevant scintillating media include but are not limited to inorganic scintillators such as NaI(Tl), CsI(Tl), and $CeBr_3$; commercial and novel plastic organic scintillators; and pulse-shape-discriminating organic scintillators such as stilbene, EJ-301, and EJ-309 for neutron background assessments.

Models of ionization quenching currently rely upon empirically derived parameters and there is currently no model of ionization quenching capable of reproducing the measured light output of scintillators over a broad energy range or for different recoil particles. Such a database is expected to improve understanding of scintillator response by providing a library with which to validate physics-based models of ionization quenching. Trends in scintillator response for classes of scintillating media may emerge to further guide theoretical studies. As the stopping power of the ionizing particle in the medium is a parameter in most extant physics-based and semi-empirical descriptions of the specific luminescence,[23,24,25,26,27] improved knowledge of the stopping power is required to test and constrain ionization quenching models for improved detector response modeling. This is particularly important for heavy ions and protons at or below the Bragg peak in novel and commercial scintillating media.

---

[18] C. Romano, et al. Proceedings of the Workshop for Applied Nuclear Data: WANDA2020, Oak Ridge National Laboratory, Technical Report No. ORNL/TM-2020/1617, 2020. https://doi.org/10.2172/1649010
[19] Amanda Lewis, Ph.D. thesis, University of California, Berkeley, 2020.
[20] Eric Matthews, Ph.D. thesis, University of California, Berkeley, 2021.
[21] D. Neudecker, et al., "Template of Expected Measurement Uncertainties" (to be published).
[22] N. Otuka, et al. "Towards a More Complete and Accurate Experimental Nuclear Reaction Data Library (EXFOR): International Collaboration Between Nuclear Reaction Data Centres (NRDC)," Nuclear Data Sheets, 120, 272 (2014). https://doi.org/10.1016/j.nds.2014.07.065
[23] J.B. Birks, Phys. Rev. 84, 364 (1951).
[24] C.N. Chou, Phys. Rev. 87, 904 (1952).
[25] J. Hong, et al. Astropart. Phys. 16, 333 (2002).
[26] S. Yoshida, et al. Nucl. Instrum. Meth. A 662, 574 (2010).
[27] R. Voltz, et al., J. Chem. Phys. 45, 3306 (1966).

## 4. DISCUSSION AND CONCLUSIONS

The session summaries and nuclear data priorities are based on a consensus of experts in the fields of reactor modeling, antineutrino spectra, detection systems, and nuclear data. Due to greatly improved computational capabilities, nuclear data can, in many cases, drive the uncertainties in a model. This combined with the need for reduced uncertainties in measurements and simulations, nuclear data requires critical examination to determine its contribution to total uncertainties. In the case of reactor antineutrinos, it was agreed that improving the nuclear data can certainly improve the confidence in the antineutrino calculations and may influence the ability to determine plutonium content in an operating reactor using antineutrino detection systems. The recommendations in this report are expected to make a marked improvement in reactor neutrino measurement uncertainties.

## 5. ACKNOWLEDGEMENTS

The authors would like to thank the presenters and session participants who contributed valuable expertise to the discussion and without whom the recommendations would not be as impactful. We also would like to thank the rapporteurs Logan Scott (ORNL), Chad Alan Lani (Penn. State), Anosh Irani (IIT), Wei-Eng Ang (TAMU), Karla Tellez-Giron-Flores (Virginia Tech), and Eric Matthews (UC Berkeley) for their hard work recording the discussions. Thank you to Lisa Felker for her logistical support.